\documentclass[11pt,english]{article}

\usepackage{hyperref,amsfonts,amsmath,amssymb,bm,a4wide,cite,subfigure,graphicx,babel}

\usepackage[T1]{fontenc}
\usepackage[latin9]{inputenc}

\numberwithin{equation}{section}

\begin{document}

\title{\textbf{Quantum field theory treatment of neutrino flavor oscillations in matter}}

\author{Maxim Dvornikov\thanks{maxim.dvornikov@gmail.com}
\\
\small{\ Pushkov Institute of Terrestrial Magnetism, Ionosphere} \\
\small{and Radiowave Propagation (IZMIRAN),} \\
\small{108840 Moscow, Troitsk, Russia}}

\date{}

\maketitle

\begin{abstract}
We study neutrino oscillations in background matter within the quantum
field theory formalism where neutrino mass eigenstates are virtual
particles. In this case, neutrino mass eigenstates are mixed owing
to the interaction with matter. Assuming that neutrinos are Majorana
particles, we find the exact propagators for massive neutrinos accounting
for the interaction with matter by solving the analog of the Dyson
equation. These propagators are used to calculate the transition probability
which coincides with the prediction of the standard quantum mechanical treatment of neutrino flavor oscillations in uniform matter.
Finally, we analyze the approximations made in our analysis.
\end{abstract}

\section{Introduction\label{sec:INTRO}}

The experimental discoveries of neutrino oscillations (see, e.g.,
Ref.~\cite{An23}) confirm that these particles are mixed and have
nonzero masses. External fields are known to affect neutrino oscillations
and even result in the new oscillations channels. It can be the case
if neutrinos have magnetic moments and interact with external electromagnetic
fields~\cite{GiuStu15}. Then, besides flavor oscillations, one has
the spin-flip, i.e. neutrino spin-flavor transitions are possible.

However, the external field influencing neutrino oscillations, more
important from the point of view of possible applications, is the
electroweak interaction with background matter. It can result in the
significant amplification of the transition probability of oscillations
known as the Mikheyev--Smirnov--Wolfenstein (MSW) effect~\cite{Wol78,MikSmi85}.
Nowadays, the MSW effect is the most plausible explanation of the
solar neutrinos deficit~\cite{FogLis04}.

Historically, neutrino flavor oscillations were described within the
quantum mechanical approach. If one refers to neutrinos in vacuum,
the neutrino mass eigenstates, which are the fundamental particles, propagate
as plane waves. Recalculating the evolution for flavor eigenstates,
one gets that the flavor content of a beam changes along the distance
of the beam propagation. The quantum mechanical approach correctly
describes neutrino oscillations in almost all reasonable situations.

Nevertheless, the quantum mechanical description has certain failures
mainly from the methodological point of view. Some of these shortcomings
are outlines in Refs.~\cite{Beu03,NauNau20}. Thus, a field theory
approach for neutrino flavor oscillations is required. Numerous attempts
to construct the theory of neutrino flavor oscillations have been
made. Some of them are reviewed in Refs.~\cite{Beu03,NauNau20}.

We mention one of these approaches based on the Quantum Field Theory
(QFT), which was proposed in Refs.~\cite{Kob82,GriSto96}. It consists
in the consideration of neutrinos, propagating from a source to a detector,
as virtual particles. Some basic details are outlined shortly in Sec.~\ref{sec:QFTFORMALISM}.
Then, this formalism was extensively studied, e.g., in Refs.~\cite{GiuKimLee93,CarChu99,AkhKop10,NauNau10,AkhWil10,KovSim24}.

Despite this QFT method links flavor oscillations and the foundations
of QFT in a closest manner, it applies to either neutrino oscillations
in vacuum or to the situation when an external field is diagonal in
the mass eigenstate basis. It happens since one uses the propagators
of neutrino mass eigenstates to compute the matrix element. In general
situation, an external field mixes different mass eigenstates. Some
exceptional cases when the interaction with external fields are diagonal
in the mass basis are considered in Refs.~\cite{Dvo11,EgoVol22}.
Thus, finding exact propagators is problematic even if one gets a
propagator for a single mass eigenstate accounting for the interaction
with an external field.

Neutrino oscillations in external fields were studied in a series
of our works~\cite{Dvo11,DvoMaa07,DvoMaa09,Dvo12}, where we considered
neutrinos as first quantized fields, i.e. we were based on the relativistic
quantum mechanics. In that method, we used the solutions of wave equations
exactly accounting for an external field. In Refs.~\cite{Dvo11,DvoMaa07,DvoMaa09,Dvo12}, we could reproduce
the results for almost all reasonable oscillations channels. However,
to get a result coinciding with the quantum mechanical approach, we
had to take a quite broad initial neutrino wave packet.

In the present work, we study neutrino flavor oscillations in matter
based on the QFT approach, i.e. we consider neutrino mass eigenstates
as virtual particles. Inspite of the difficulties mentioned above,
we manage to account for the interaction of the mass eigenstates with
matter exactly. Thus, based on reasonable approximations, we find
the exact propagators of the mass eigenstates in matter. Eventually, using our results,
we reproduce the transition probability for neutrino flavor oscillations in uniform matter obtained within the quantum mechanical approach.

This work is organized in the following way. We start in Sec.~\ref{sec:QFTFORMALISM}
with a brief description of the formalism for neutrino flavor oscillations
within QFT. Then, in Sec.~\ref{sec:NUMATT}, we remind how mixed
neutrinos interact with background matter. In Sec.~\ref{sec:EXPROP},
considering two flavor case, we derive the equations for the exact
propagators of massive neutrinos in background matter. The developed
formalism is adapted for Majorana neutrinos in Sec.~\ref{sec:REFORMMAJ}.
The exact propagators for massive Weyl neutrinos in matter are found
in Sec.~\ref{sec:PROPWEYLNU}. These propagators are used in Sec.~\ref{sec:MSW}
to find the transition probability for neutrino flavor oscillations
in matter within the QFT approach. The detailed comparison of QFT and quantum mechanical approaches is provided in Sec.~\ref{sec:QMvsQFT}. Finally, we conclude in Sec.~\ref{sec:CONCL}.
The propagators of massive Weyl neutrinos accounting for the diagonal
interaction with matter are derived in Appendix~\ref{sec:MATTPROP}.
In Appendix~\ref{sec:FOURTRANS}, we compute the specific 3D Fourier
transforms of the propagators obtained in Appendix~\ref{sec:MATTPROP}.

\section{Quantum field theory formalism for neutrino flavor oscillations\label{sec:QFTFORMALISM}}

In this section, we outline the main features of the description of
flavor oscillations where neutrinos are virtual particles. This formalism
was proposed first in Ref.~\cite{Kob82} and then independently studied
in Ref.~\cite{GriSto96}.

\begin{figure}
  \centering
  \includegraphics[viewport=100bp 600bp 350bp 750bp,clip,scale=0.7]{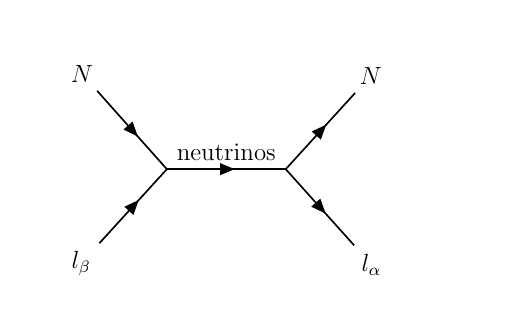}
  \protect
  \caption{The schematic illustration of neutrino oscillations in QFT.\label{fig:feynQFT}}
\end{figure}

We suppose that a source and a detector of neutrinos are separated
in space by the distance $\mathbf{L}$. The neutrino beam in a source
is created by the interaction of charged leptons belonging to a certain
flavor $\beta$ with the heavy nucleus $N$. The neutrino detection
is implemented again by the interaction of charged leptons of the flavor
$\alpha$ the a nucleus. If $\alpha\neq\beta$, one can speak about
neutrino oscillations. The schematic Feynman diagram for the described
process is depicted in Fig.~\ref{fig:feynQFT}.

The interaction between leptons and nuclei is described by the Lagrangian,
\begin{equation}\label{eq:nucllepint}
  \mathcal{L}_{\mathrm{int}}=\sqrt{2}G_{\mathrm{F}}j_{\mu}J^{\mu},
\end{equation}
where $G_{\mathrm{F}}$ is the coupling constant, e.g., it can be
the Fermi constant,
\begin{equation}\label{eq:lepcurr}
  j_{\mu}=\sum_{\lambda}\bar{\nu}_{\lambda\mathrm{L}}\gamma_{\mu}l_{\lambda\mathrm{L}},
\end{equation}
is the leptonic current, $\gamma^{\mu}=(\gamma^{0},\bm{\gamma})$
are the Dirac matrices, $\nu_{\lambda}$ is the neutrino wavefunction
corresponding to a certain flavor $\lambda=(e,\mu,\tau)$, $l_{\lambda}$
is the wavefunction of the charged lepton, and $J^{\mu}$ is the nuclear
current. In Eq.~(\ref{eq:lepcurr}), we assume that only left chiral
projections of neutrinos and leptons participate in the interaction.
For instance, $\nu_{\lambda\mathrm{L}}=P_{\mathrm{L}}\nu_{\lambda}$,
where $P_{\mathrm{L}}=(1-\gamma^{5})/2$ and $\gamma^{5}=\mathrm{i}\gamma^{0}\gamma^{1}\gamma^{2}\gamma^{3}$
is the Dirac matrix.

Based on Eq.~(\ref{eq:nucllepint}), one gets the $S$-matrix element
for the process $(l_{\beta}N)\to\text{neutrinos}\to(l_{\alpha}N)$
in the form,
\begin{equation}\label{eq:Smatrgen}
  S=-\frac{1}{2}
  \left(
    \sqrt{2}G_\mathrm{F}
  \right)^{2}
  \int\mathrm{d}^{4}x\mathrm{d}^{4}y
  \left\langle
    l_{\alpha}
    \left|
      T
      \left\{
         j_{\mu}^{\dagger}(x)J^{\mu}(x)j^{\nu}(y)J_{\nu}^{\dagger}(y)
      \right\}
    \right|
    l_{\beta}
  \right\rangle,
\end{equation}
where $T$ stays for the chronological ordering of operators.

If charged leptons have the flavors $\beta$ in a source and $\alpha$
in a detector, then the corresponding operators in Eq.~(\ref{eq:lepcurr})
act as
\begin{equation}\label{eq:lepoper}
  \hat{l}_{\lambda}(y)
  \left|
    l_{\beta}
  \right\rangle =
  \frac{\delta_{\lambda\beta}}{\sqrt{2VE_{\beta}}}e^{-\mathrm{i}p_{\beta}y}u(p_{\beta})
  \left| 0 \right\rangle,
  \quad
  \left\langle
    l_{\alpha}
  \right|
  \hat{\bar{l}}_{\lambda}(x)=\frac{\delta_{\lambda\beta}}{\sqrt{2VE_{\alpha}}}e^{\mathrm{i}p_{\alpha}x}\bar{u}(p_{\alpha})
  \left\langle 0 \right|,
\end{equation}
where $V$ is the normalization volume, $p_{\alpha,\beta}^{\mu}=(E_{\alpha,\beta},\mathbf{p}_{\alpha,\beta})$
are the four-momenta of outgoing and incoming leptons, and $u(p_{\alpha,\beta})$
are their $c$-number wavefunctions. One implies in Eq.~(\ref{eq:lepoper})
that leptons propagate as plane waves.

The target nuclei are supposed to be quite heavy. Thus, their currents
can be written in the form,
\begin{equation}\label{eq:nuclcurr}
  J_{\mu}(x)\propto\delta_{\mu0}\delta(\mathbf{x}-\mathbf{x}_{2}),\quad J_{\nu}(y)\propto\delta_{\nu0}\delta(\mathbf{y}-\mathbf{x}_{1}),
\end{equation}
where $\mathbf{x}_{1,2}$ are the positions of the source and the
detector, i.e. $\mathbf{L}=\mathbf{x}_{2}-\mathbf{x}_{1}$.

Flavor neutrinos $\nu_{\lambda}$ in Eq.~(\ref{eq:lepcurr}) is the
superposition of the mass eigenstates $\psi_{a}$,
\begin{equation}\label{eq:nuflmass}
  \nu_{\lambda}=\sum_{a}U_{\lambda a}\psi_{a},
\end{equation}
where $(U_{\lambda a})$ is the mixing matrix. If we consider only
active neutrinos, $(U_{\lambda a})$ is a $3\times3$ unitary matrix.

Using Eqs.~(\ref{eq:lepoper})-(\ref{eq:nuflmass}), we rewrite Eq.~(\ref{eq:Smatrgen})
as
\begin{align}\label{eq:Smatrx0y0}
  S= & -\frac{G_{\mathrm{F}}^{2}e^{-\mathrm{i}\mathbf{p}_{\alpha}\mathbf{x}_{2}+\mathrm{i}\mathbf{p}_{\beta}\mathbf{x}_{1}}}
  {2V\sqrt{E_{\alpha}E_{\beta}}}  
  \sum_{ab}U_{\alpha a}U_{\beta b}^{*}\int\mathrm{d}x_{0}\mathrm{d}y_{0}e^{\mathrm{i}E_{\alpha}x_{0}-\mathrm{i}E_{\beta}y_{0}}
  \nonumber
  \\
  & \times
  \bar{u}_{\mathrm{L}}(p_{\alpha})\gamma^{0}
  \left\langle 0\left|
    T
    \left\{
      \psi_{a\mathrm{L}}(x_{0},\mathbf{x}_{2})\bar{\psi}_{b\mathrm{L}}(y_{0},\mathbf{x}_{1})
    \right\}
  \right|0\right\rangle
  \gamma^{0}u_{\mathrm{L}}(p_{\beta}).
\end{align}
Based on the Dirac matrices in the chiral representation, one transforms
the factor in the integrand in Eq.~(\ref{eq:Smatrx0y0}),
\begin{equation}\label{eq:preprop}
  \bar{u}_{\mathrm{L}}(p_{\alpha})\gamma^{0}
  \left\langle 0\left|
    T
    \left\{
      \psi_{a\mathrm{L}}(x)\bar{\psi}_{b\mathrm{L}}(y)
    \right\}
  \right|0\right\rangle
  \gamma^{0}u_{\mathrm{L}}(p_{\beta})=\kappa_{-}^{\dagger}(p_{\alpha})
  \left\langle 0\left|
    T
    \left\{
      \eta_{a}(x)\eta_{b}^{\dagger}(y)
    \right\}
  \right|0\right\rangle
  \kappa_{-}(p_{\beta}),
\end{equation}
where $\eta_{a,b}$ and $\kappa(p_{\alpha,\beta})$ are the two component
spinors defined by $\psi_{a,b\mathrm{L}}^{\mathrm{T}}=(0,\eta_{a,b})$
and $u_{\mathrm{L}}^{\mathrm{T}}(p_{\alpha,\beta})=(0,\kappa(p_{\alpha,\beta}))$.
In Eq.~(\ref{eq:preprop}), we assume that the charged leptons are
ultrarelativistic, $E_{\alpha,\beta}\gg m_{\alpha,\beta}$. In this
case, only $\kappa_{-}(p_{\alpha,\beta})=-\sqrt{2E_{\alpha,\beta}}w_{-}\neq0$,
where $w_{-}$ is the helicity amplitude given in Eq.~(\ref{eq:helamp}).

If we consider massive Majorana neutrinos, we can represent them as
Weyl fields $\eta_{a}$. In this situation, the expressions $\left\langle 0\left|T\left\{ \eta_{a}(x)\eta_{b}^{\dagger}(y)\right\} \right|0\right\rangle =\Sigma_{ab}(x-y)$
are the propagators. We discuss them in Sec.~\ref{sec:EXPROP}.

The remaining integrals in Eq.~(\ref{eq:Smatrx0y0}) can be calculated
using the new variables $t=x_{0}-y_{0}$ and $T=(x_{0}+y_{0})/2$.
The final result has the form,
\begin{equation}\label{eq:Smatrftprop}
  S=-\frac{G_{\mathrm{F}}^{2}e^{-\mathrm{i}\mathbf{p}_{\alpha}\mathbf{x}_{2}+\mathrm{i}\mathbf{p}_{\beta}\mathbf{x}_{1}}}
  {2V\sqrt{E_{\alpha}E_{\beta}}}
  2\pi\delta(E_{\alpha}-E_{\beta})\mathcal{M}_{\beta\to\alpha},
\end{equation}
where the matrix element is
\begin{equation}\label{eq:matrel}
  \mathcal{M}_{\beta\to\alpha}=
  \sum_{ab}U_{\alpha a}U_{\beta b}^{*}
  \kappa_{-}^{\dagger}(p_{\alpha})
  \left(
    \int\frac{\mathrm{d}^{3}q}{(2\pi)^{3}}\Sigma_{ab}(E,\mathbf{q})e^{\mathrm{i}\mathbf{qL}}
  \right)
  \kappa_{-}(p_{\beta}).
\end{equation}
In Eq.~\eqref{eq:matrel}, $\Sigma_{ab}(p)$ is the Fourier transform
of $\Sigma_{ab}(x)$ and $E=(E_{\alpha}+E_{\beta})/2$.

The probability of the transition $\beta\to\alpha$ is $P_{\beta\to\alpha}\propto|\mathcal{M}_{\beta\to\alpha}|^{2}$.
If virtual massive neutrinos propagate in vacuum, the propagators
are diagonal, $\Sigma_{ab}\propto\delta_{ab}$. In this situation,
the calculation of $P_{\beta\to\alpha}$ is quite straightforward;
cf. Refs.~\cite{Kob82,GriSto96}. It gives one the expression for
the probabilities of flavor neutrinos oscillations in vacuum. Our
main goal is to find $\Sigma_{ab}$ and $\mathcal{M}_{\beta\to\alpha}$
for neutrinos propagating in background matter.

\section{Mixed neutrinos in background matter\label{sec:NUMATT}}

In this section, we briefly recall how neutrinos interact with background
matter in frames of the field theory.

For simplicity, we consider the system of two mixed active flavor
neutrinos $\nu_{\lambda}$, e.g., $\lambda=e,\mu$, in the presence of
background matter. This matter is supposed to be electroneutral, nonmoving,
and unpolarized. The neutrino interaction with matter is treated in
the forward scattering approximation. In this case, the Lagrangian
for this system has the form,
\begin{equation}\label{eq:Lagrflavor}
  \mathcal{L}=\sum_{\lambda\lambda'}\bar{\nu}_{\lambda}
  \left[
    \delta_{\lambda\lambda'}\mathrm{i}\gamma^{\mu}\partial_{\mu}-m_{\lambda\lambda'}
    -V_{\lambda\lambda'}\frac{1}{2}\gamma^{0}(1-\gamma^{5})
  \right]
  \nu_{\lambda'},
\end{equation}
where $(m_{\lambda\lambda'})$ is the nondiagonal neutrino mass matrix.
The matrix of the effective potentials for the neutrino interaction
with matter $(V_{\lambda\lambda'})$ is diagonal in the flavor basis
in case of standard neutrino interactions, $V_{\lambda\lambda'}=\delta_{\lambda\lambda'}V_{\lambda}$.
For example, 
\begin{equation}\label{eq:VeVmu}
  V_{e}=\sqrt{2}G_{\mathrm{F}}\left(n_{e}-\frac{1}{2}n_{n}\right),
  \quad
  V_{\mu}=-\frac{G_{\mathrm{F}}}{\sqrt{2}}n_{n},
\end{equation}
for two neutrinos, $(\nu_{e},\nu_{\mu})$, in matter consisting of
electrons, protons, and neutrons. In Eq.~(\ref{eq:VeVmu}), $n_{e}=n_{p}$
and $n_{n}$ are the number densities of background fermions, and
$G_{\mathrm{F}}=1.17\times10^{-5}\,\text{GeV}^{-2}$ is the Fermi
constant.

Then, we diagonalize the mass matrix and introduce the mass eigenstates,
$\psi_{a}$, by means of the matrix transformation in Eq.~\eqref{eq:nuflmass},
$\text{diag}(m_{1},m_{2})=U^{\dagger}(m_{\lambda\lambda'})U$, where
$m_{1,2}$ are the masses of $\psi_{1,2}$. In case of two neutrinos,
$(U_{\lambda a})$ is represented in the form,
\begin{equation}\label{eq:2flmixmatr}
  (U_{\lambda a})=
  \left(
    \begin{array}{cc}
      \cos\theta & \sin\theta
      \\
      -\sin\theta & \cos\theta
    \end{array}
  \right),
\end{equation}
where $\theta$ is the vacuum mixing angle.

The Lagrangian in Eq.~(\ref{eq:Lagrflavor}), rewritten in terms
of the mass eigenstates, takes the form,
\begin{equation}\label{eq:Lagrmass}
  \mathcal{L}=\sum_{ab}\bar{\psi}_{a}
  \left[
    \delta_{ab}\left(\mathrm{i}\gamma^{\mu}\partial_{\mu}-m_{a}\right)-g_{ab}\frac{1}{2}\gamma^{0}(1-\gamma^{5})
  \right]
  \psi_{b},
\end{equation}
where the matrix of effective potentials $(g_{ab})=U^{\dagger}VU$
is nondiagonal in general case. The Lagrangian in Eq.~(\ref{eq:Lagrmass})
is valid for both Dirac and Majorana neutrinos $\psi_{a}$. In Majorana
case, we just replace $\frac{1}{2}\gamma^{0}(1-\gamma^{5})\to-\gamma^{0}\gamma^{5}$.

The wave equations for $\psi_{a}$, $a=1,2$, resulting from Eq.~(\ref{eq:Lagrmass}),
have the form,
\begin{align}\label{eq:psi12waveeq}
  \left[
    \mathrm{i}\gamma^{\mu}\partial_{\mu}-m_{1}-\frac{g_{1}}{2}\gamma^{0}(1-\gamma^{5})
  \right]
  \psi_{1} & =
  \frac{g}{2}\gamma^{0}(1-\gamma^{5})\psi_{2},
  \nonumber
  \\
  \left[
    \mathrm{i}\gamma^{\mu}\partial_{\mu}-m_{2}-\frac{g_{2}}{2}\gamma^{0}(1-\gamma^{5})
  \right]
  \psi_{2} & =
  \frac{g}{2}\gamma^{0}(1-\gamma^{5})\psi_{1},
\end{align}
where $g\equiv g_{12}\neq0$. One can see in Eq.~(\ref{eq:psi12waveeq})
that the mass eigenstates are converted into each other in background
matter.

\section{Exact propagators of massive neutrinos in matter\label{sec:EXPROP}}

In this section, we show how to obtain the exact propagators of neutrino
mass eigenstates $\Sigma_{ab}$, which were defined in Sec.~\ref{sec:QFTFORMALISM},
in background matter. Thus, we consider a situation when the space
between the source and the detector is filled with matter which a neutrino
interacts with.

If a field $\psi$ obeys a wave equation $\mathcal{D}\psi=0$, where
$\mathcal{D}$ is the differential operator, the Green function, e.g.,
a propagator $S$, can be found as the solution of the equation, $\mathcal{D}S=\delta(x)$,
with the same differential operator. Bypassing the poles in the formal
solution, one gets a Green function with required properties.

One can see in Eqs.~(\ref{eq:psi12waveeq}) that it is problematic
to solve the equation for a Green function even for two neutrinos.
Indeed, one should consider the multiplet $\Psi=(\psi_{1},\psi_{2})$
and has to invert the $8\times8$ matrix in Dirac indices while looking
for a Green function. Nevertheless, we can tackle the problem of finding
the propagators of mass eigenstates perturbatively and try to sum
all the terms in a perturbation series.

We define the diagonal propagators of mass eigenstates as $S_{a}$,
which obey the equations,
\begin{equation}\label{eq:SaeqDir}
  \left[
    \mathrm{i}\gamma^{\mu}\partial_{\mu}-m_{a}-\frac{g_{a}}{2}\gamma^{0}(1-\gamma^{5})
  \right]
  S_{a}=\delta(x).
\end{equation}
These propagators exactly account for the diagonal interactions with
matter $g_{a}$. In case of Dirac neutrinos, $S_{a}$ were found in
Ref.~\cite{Stu08} (see also Ref.~\cite{DvoSem14}). The propagators
in Eq.~(\ref{eq:SaeqDir}) for Majorana neutrinos are obtained in
Appendix~\ref{sec:MATTPROP}. 

\begin{figure}[htbp]
  \centering
  \subfigure[]
  {\label{fig:f1a}
  \includegraphics[viewport=150bp 600bp 450bp 730bp,clip,scale=0.65]{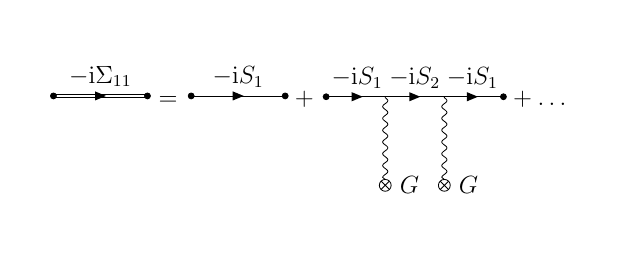}}
  \subfigure[]
  {\label{fig:f1b}
  \includegraphics[viewport=150bp 600bp 450bp 730bp,clip,scale=0.65]{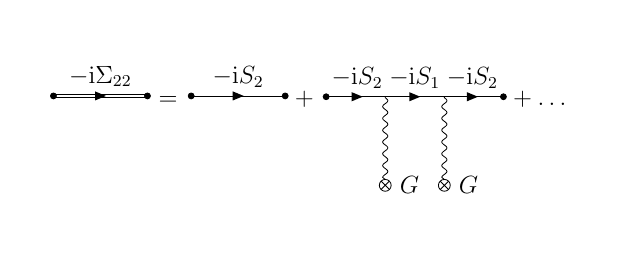}}
  \\
  \subfigure[]
  {\label{fig:f1c}
  \includegraphics[viewport=150bp 600bp 450bp 730bp,clip,scale=0.65]{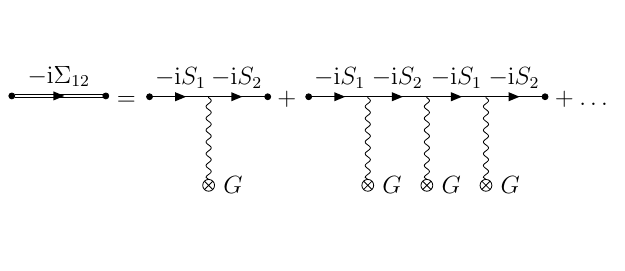}}
  \subfigure[]
  {\label{fig:f1d}
  \includegraphics[viewport=150bp 600bp 450bp 730bp,clip,scale=0.65]{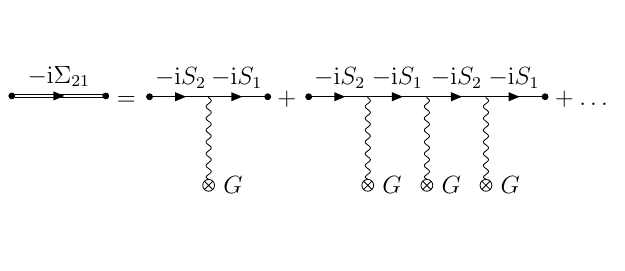}}
  \protect 
\caption{Feynman diagrams corresponding to Eqs.~(\ref{eq:Sigmaaser}) and~(\ref{eq:Sigma12ser}).
4D Fourier images of the bare propagators $-\mathrm{i}S_{a}$ are
depicted by thin solid lines, whereas thick solid lines correspond
to the dressed propagators $-\mathrm{i}\Sigma_{ab}$~\cite{BerLifPit82}.\label{fig:DysonFeyn}}
\end{figure}

To account for the transitions $1\leftrightarrow2$ in matter, we
notice that they are induced by the nondiagonal potential $G=\frac{g}{2}\gamma^{0}(1-\gamma^{5})$.
For example, the exact propagators for the processes $1\to1$ and
$2\to2$, $\Sigma_{11}$ and $\Sigma_{22}$, are the sums of the formal
series,
\begin{align}\label{eq:Sigmaaser}
  -\mathrm{i}\Sigma_{11} & =-\mathrm{i}S_{1}+(-\mathrm{i}S_{1})G(-\mathrm{i}S_{2})G(-\mathrm{i}S_{1})+\dotsb,
  \nonumber
  \\
  -\mathrm{i}\Sigma_{22} & =-\mathrm{i}S_{2}+(-\mathrm{i}S_{2})G(-\mathrm{i}S_{1})G(-\mathrm{i}S_{2})+\dotsb,
\end{align}
whereas for the processes $1\to2$ and $2\to1$, $\Sigma_{12}$ and
$\Sigma_{21}$, satisfy the expressions,
\begin{align}\label{eq:Sigma12ser}
  -\mathrm{i}\Sigma_{12} & =
  (-\mathrm{i}S_{1})G(-\mathrm{i}S_{2})+(-\mathrm{i}S_{1})G(-\mathrm{i}S_{2})G(-\mathrm{i}S_{1})G(-\mathrm{i}S_{2})+\dotsb,  
  \nonumber
  \\
  -\mathrm{i}\Sigma_{21} & =
  (-\mathrm{i}S_{2})G(-\mathrm{i}S_{1})+(-\mathrm{i}S_{2})G(-\mathrm{i}S_{1})G(-\mathrm{i}S_{2})G(-\mathrm{i}S_{1})+\dotsb.
\end{align}
In Eqs.~(\ref{eq:Sigmaaser}) and~(\ref{eq:Sigma12ser}), both $S_{1,2} = S_{1,2}(p)$ and $\Sigma_{ab}$ are the 4D Fourier images of the propagators $S_{1,2} = S_{1,2}(x)$ and $\Sigma_{ab}(x)$ respectively. Note that the treatment of neutrino
oscillations in matter analogous to that in Eqs.~(\ref{eq:Sigmaaser})
and~(\ref{eq:Sigma12ser}) was made in Ref.~\cite{Tur23}. The analytical
expressions in Eqs.~(\ref{eq:Sigmaaser}) and~(\ref{eq:Sigma12ser})
are illustrated in Fig.~\ref{fig:DysonFeyn}.

The sum of the series in Eqs.~(\ref{eq:Sigmaaser}) and~(\ref{eq:Sigma12ser})
can be represented in the form of the Dyson equation for an exact
Green function,
\begin{align}\label{eq:DysonSigmaa}
  (-\mathrm{i}S_{1})^{-1} & =(-\mathrm{i}\Sigma_{11})^{-1}+G(-\mathrm{i}S_{2})G,
  \nonumber
  \\
  (-\mathrm{i}S_{2})^{-1} & =(-\mathrm{i}\Sigma_{22})^{-1}+G(-\mathrm{i}S_{1})G,
\end{align}
and
\begin{align}\label{eq:DysonSigma12}
  \left[
    (-\mathrm{i}S_{1})G(-\mathrm{i}S_{2})
  \right]^{-1} & =
  (-\mathrm{i}\Sigma_{12})^{-1}+G,
  \nonumber
  \\
  \left[
    (-\mathrm{i}S_{2})G(-\mathrm{i}S_{1})
  \right]^{-1} & =
  (-\mathrm{i}\Sigma_{21})^{-1}+G.
\end{align}
The diagonal propagators $S_{1,2}$, which are in Eq.~\eqref{eq:DysonSigmaa}, were found in Ref.~\cite{Stu08} for the case of Dirac neutrinos. They are $4\times 4$ matrices with the quite nontrivial Dirac matrices structure. Thus, despite the solutions of Eq.~(\ref{eq:DysonSigmaa}), in principle, exist for Dirac neutrinos, finding them may be rather complicated in practice.
One can also notice that Eq.~(\ref{eq:DysonSigma12}) implies the inversion
of the effective potential $G$. However, it contains the projection
$(1-\gamma^{5})/2$ for Dirac neutrinos. It means that $G^{-1}$ does
not exist in this situation. Therefore, the treatment of neutrino flavor oscillations, proposed  in our work, cannot be used for Dirac neutrinos.

Nevertheless, the developed formalism is well defined for
Majorana particles. It was found, e.g., in Refs.~\cite{Kob80,SchVal80} that the diagonalization of the neutrino mass matrix of the general form, which includes both left and right Majorana terms, as well as the Dirac term, gives one certain number of Majorana mass eigenstates. Moreover, the most natural  explanation of the small masses of active neutrinos implies that mass eigenstates are Majorana particles (see, e.g., Ref.~\cite{MirVal16}). It is the heuristic argument for considering Majorana neutrinos in our work.

It is known (see, e.g., Ref.~\cite{GiuKim07}) that a Dirac mass eigenstate can be treated as two Majorana neutrinos with equal masses. However, such Dirac neutrinos cannot be studied in frames of our approach since it requires the consideration of more than two types of particles. As one can see in Eq.~\eqref{eq:psi12waveeq}, it is essential in our method that one deals with only two massive degrees of freedom. We also analyze this approximation in Sec.~\ref{sec:CONCL}.

\section{Reformulation of the formalism for Majorana mass eigenstates\label{sec:REFORMMAJ}}

In this section, we rewrite the basic expressions in Secs.\ref{sec:NUMATT}
and~\ref{sec:EXPROP} for Majorana neutrinos.

Since we found in Sec.~\ref{sec:EXPROP} that transitions between
mass eigenstates in matter can be described for Majorana neutrinos,
we should rewrite the basic expressions for this kind of particles.
Instead of dealing with a four components wave function, we use the
Weyl representation for a Majorana neutrino in terms of the two component
spinor $\eta$. It is made in order to avoid the constraint, i.e.
the Majorana condition, in the quantization of the fields. If we use
the chiral representation of Dirac matrices, the four component Majorana
wave function has the form $\psi^{\mathrm{T}}=(\mathrm{i}\sigma_{2}\eta^{*},\eta)$,
where $\eta$ corresponds to a left-handed Weyl neutrino, and $\sigma_{2}$
is the Pauli matrix.

The interaction of a Weyl neutrino mass eigenstate with matter takes
the form,
\begin{equation}
  \mathcal{L}_{\mathrm{int}}=\sum_{ab}g_{ab}\bar{\psi}_{a\mathrm{L}}\gamma^{0}\gamma^{5}\psi_{b\mathrm{L}}=
  -\sum_{ab}g_{ab}\eta_{a}^{\dagger}\eta_{b}.
\end{equation}
The analog of Eq.~(\ref{eq:psi12waveeq}) for Weyl neutrinos is (see
also Appendix~\ref{sec:MATTPROP})
\begin{align}\label{eq:eta12waveeq-1}
  \mathrm{i}\dot{\eta}_{1}+[(\bm{\sigma}\mathbf{p})-g_{1}]\eta_{1}+\mathrm{i}m_{1}\sigma_{2}\eta_{1}^{*} & =g\eta_{2},
  \nonumber
  \\
  \mathrm{i}\dot{\eta}_{2}+[(\bm{\sigma}\mathbf{p})-g_{2}]\eta_{2}+\mathrm{i}m_{2}\sigma_{2}\eta_{2}^{*} & =g\eta_{1},
\end{align}
where $\bm{\sigma}=(\sigma_{1},\sigma_{2},\sigma_{3})$ are the Pauli
matrices, $\mathbf{p}=-\mathrm{i}\nabla$, and a dot means the time
derivative.

The relation between $(g_{1,2},g)$ and $V_{e,\mu}$ is the same as
for Dirac neutrinos. The main results of Sec.~\ref{sec:EXPROP},
e.g., Eqs.~(\ref{eq:DysonSigmaa}) and~(\ref{eq:DysonSigma12}),
remain valid for Weyl neutrinos as well. However, $G=g$ is the scalar
quantity.

The propagators of Weyl neutrinos $S_{a}$, accounting for the diagonal
interaction with matter $g_{a}$, are derived in Appendix~\ref{sec:MATTPROP};
cf. Eq.~(\ref{eq:Sprop}). In our analysis, we consider only the transformations
of particles in background matter. It means that, while studying,
e.g., the process $1\to2$, we neglect the chain like $(1\to\bar{1})\xrightarrow{g}(\bar{2}\to2)$,
where the blocks $1\to\bar{1}$ and $\bar{2}\to2$ stand for neutrino-to-antineutrino
oscillations, and the symbol $g$ over the arrow means the transition $\bar{1}\to\bar{2}$
induced by the matter interaction. Such a term implies using the propagator
$\tilde{S}_{a}$ in Eq.~(\ref{eq:tildeSprop}). As shown in Appendix~\ref{sec:MATTPROP}
(see also Ref.~\cite{Kob82}), the transitions like $a\to\bar{a}$
are suppressed by the factor $m_{a}/E$, where $E$ is the typical
neutrino energy. Thus, we can safely neglect them if we study ultrarelativistic
particles.

\section{Exact propagators of massive Weyl neutrinos\label{sec:PROPWEYLNU}}

In this section, we solve Eqs.~\eqref{eq:DysonSigmaa} and~\eqref{eq:DysonSigma12}
for Majorana neutrinos.

We represent the 4D Fourier image of the propagator $S_{a}$ in Eq.~(\ref{eq:Sprop})
in the form,
\begin{equation}\label{eq:Sa2fl}
  S_{a}(p)=\mathcal{A}_{a}(p)+(\bm{\sigma}\hat{p})\mathcal{B}_{a}(p),
\end{equation}
where both $\mathcal{A}_{a}$ and $\mathcal{B}_{a}$ contain four
terms. Then, after tedious but straightforward calculations, one finds
the solution of Eqs.~(\ref{eq:DysonSigmaa}) and~(\ref{eq:DysonSigma12}),
\begin{align}\label{eq:SigmaAB}
  \Sigma_{11} & =
  \frac{\mathcal{A}_{1}+(\bm{\sigma}\hat{p})\mathcal{B}_{1}
  +g^{2}(\mathcal{A}_{1}^{2}-\mathcal{B}_{1}^{2})(\mathcal{A}_{2}-(\bm{\sigma}\hat{p})\mathcal{B}_{2})}
  {(1+g^{2}\mathcal{A}_{1}\mathcal{A}_{2})^{2}+(1+g^{2}\mathcal{B}_{1}\mathcal{B}_{2})^{2}
  -1-g^{4}(\mathcal{A}_{1}^{2}\mathcal{B}_{2}^{2}+\mathcal{A}_{2}^{2}\mathcal{B}_{1}^{2})},
  \nonumber
  \\
  \Sigma_{22} & =
  \frac{\mathcal{A}_{2}+(\bm{\sigma}\hat{p})\mathcal{B}_{2}
  +g^{2}(\mathcal{A}_{2}^{2}-\mathcal{B}_{2}^{2})(\mathcal{A}_{1}-(\bm{\sigma}\hat{p})\mathcal{B}_{1})}
  {(1+g^{2}\mathcal{A}_{1}A_{2})^{2}+(1+g^{2}B_{1}B_{2})^{2}
  -1-g^{4}(\mathcal{A}_{1}^{2}B_{2}^{2}+\mathcal{A}_{2}^{2}B_{1}^{2})},
  \nonumber
  \\
  \Sigma_{12} & = \Sigma_{21}=
  \frac{\mathrm{i}g
  \left[
    \mathcal{A}_{1}\mathcal{A}_{2}+\mathcal{B}_{1}\mathcal{B}_{2}
    -g^{2}(\mathcal{A}_{1}^{2}-\mathcal{B}_{1}^{2})(\mathcal{A}_{2}^{2}-\mathcal{B}_{2}^{2})
    -(\bm{\sigma}\hat{p})(\mathcal{A}_{1}\mathcal{B}_{2}-\mathcal{B}_{1}\mathcal{A}_{2})
  \right]}
  {(1+g^{2}\mathcal{A}_{1}\mathcal{A}_{2})^{2}+(1+g^{2}\mathcal{B}_{1}\mathcal{B}_{2})^{2}
  -1-g^{4}(\mathcal{A}_{1}^{2}\mathcal{B}_{2}^{2}+\mathcal{A}_{2}^{2}\mathcal{B}_{1}^{2})},
\end{align}
which is valid for neutrinos with arbitrary energies.

Further analysis of Eq.~(\ref{eq:SigmaAB}) is quite difficult since
$\mathcal{A}_{a}$ and $\mathcal{B}_{a}$ have the complicated structure.
However, as shown in Appendix~\ref{sec:FOURTRANS}, the main contribution
to the matrix element involving the 3D Fourier transform in Eq.~\eqref{eq:ftprop}
comes from $I_{1}$ and $J_{1}$ terms. That is why, we keep only
one term in $\mathcal{A}_{a}$ and $\mathcal{B}_{a}$, i.e.
\begin{equation}\label{eq:ABultrarel}
  \mathcal{A}_{a}=-\mathcal{B}_{a}=\frac{\mathrm{i}\lambda_{a-}^{2}}{2(p_{0}-E_{a-}+\mathrm{i}0)}.
\end{equation}
Note that Eq.~(\ref{eq:ABultrarel}) is valid for ultrarelativistic
neutrinos.

To proceed, we assume that the interaction with matter is weak. In
this approximation, we keep the terms up to $\sim g^{2}$ in Eq.~(\ref{eq:SigmaAB}).
Using Eq.~(\ref{eq:ABultrarel}), we rewrite Eq.~(\ref{eq:SigmaAB})
in the form,
\begin{align}\label{eq:Sigmaab2fl}
  \Sigma_{11}(p_{0},\mathbf{p}) & =
  \frac{\mathrm{i}[1-(\bm{\sigma}\hat{p})](p_{0}-E_{2-})}{2[(p_{0}-E_{1-}+\mathrm{i}0)(p_{0}-E_{2-}+\mathrm{i}0)-g^{2}]},
  \nonumber
  \\
  \Sigma_{22}(p_{0},\mathbf{p}) & =
  \frac{\mathrm{i}[1-(\bm{\sigma}\hat{p})](p_{0}-E_{1-})}{2[(p_{0}-E_{1-}+\mathrm{i}0)(p_{0}-E_{2-}+\mathrm{i}0)-g^{2}]},
  \nonumber
  \\
  \Sigma_{12}(p_{0},\mathbf{p}) & =
  \frac{\mathrm{i}g[1-(\bm{\sigma}\hat{p})]}{2[(p_{0}-E_{1-}+\mathrm{i}0)(p_{0}-E_{2-}+\mathrm{i}0)-g^{2}]}.
\end{align}
The exact propagators $\Sigma_{ab}$, accounting for the interaction
of massive neutrinos with matter, are used in Eq.~(\ref{eq:matrel})
to calculate the matrix element.

\section{Transition probability for oscillations in matter\label{sec:MSW}}

In this section, we apply our results to derive the transition probability
of neutrino oscillations in matter basing on QFT.

Let us consider the particular oscillations channel, $e\to\mu$. Using
Eqs.~(\ref{eq:matrel}) and~(\ref{eq:2flmixmatr}), we get that
the matrix element is
\begin{equation}\label{eq:Memu}
  \mathcal{M}_{e\to\mu}\propto\kappa_{-}^{\dagger}(p_{\mu})
  \left[
    \sin2\theta\frac{1}{2}(\Sigma_{22}-\Sigma_{11})+\cos2\theta\Sigma_{12}
  \right]
  \kappa_{-}(p_{e}),
\end{equation}
where the exact propagators $\Sigma_{ab}(p_{0},\mathbf{p})$ are given
in Eq.~(\ref{eq:Sigmaab2fl}). To get the final form of the matrix
element in Eq.~(\ref{eq:Memu}) one should put $p_{0}=E$, calculate
the 3D Fourier transform, as well as average the result over the incoming
and outgoing left-handed charged leptons.

We make the above calculations in details, e.g., for the $\Sigma_{12}$
contribution. After substituting $p_{0}\to E$, we transform the factor
in the integrand
\begin{equation}\label{eq:demondecomp}
  \frac{1}{(E-E_{1-}+\mathrm{i}0)(E-E_{2-}+\mathrm{i}0)-g^{2}}=
  \frac{1}{\mathcal{E}_{+}-\mathcal{E}_{-}}
  \left(
    \frac{1}{E-\mathcal{E}_{+}+\mathrm{i}0}-\frac{1}{E-\mathcal{E}_{-}+\mathrm{i}0}
  \right),
\end{equation}
where
\begin{equation}\label{eq:Epm}
  \mathcal{E}_{\pm}=\frac{E_{1-}+E_{2-}}{2}\pm\sqrt{
  \left(
    \frac{E_{1-}-E_{2-}}{2}
  \right)^{2}
  +g^{2}}.
\end{equation}
Note that $\mathcal{E}_{\pm}$ depend quite non-trivially on $q$.

Based on Eqs.~(\ref{eq:demondecomp}) and~(\ref{eq:Epm}), we compute,
e.g., the integral
\begin{equation}
  I=\int\frac{\mathrm{d}^{3}q}{(2\pi)^{3}}\frac{e^{\mathrm{i}\mathbf{qL}}}{(\mathcal{E}_{+}-\mathcal{E}_{-})(E-\mathcal{E}_{+}+\mathrm{i}0)},
\end{equation}
which contributes to the 3D Fourier transform of $\Sigma_{12}$. Analogously
to Appendix~\ref{sec:FOURTRANS}, we use the cylindrical coordinates
$\mathbf{q}=(\rho,z,\phi)$ and assume that $\mathbf{L}$ is along
the positive direction of $\mathbf{e}_{z}$. Then,
\begin{equation}
  I=-\frac{\mathrm{i}}{2\pi}\int_{0}^{+\infty}\rho\mathrm{d}\rho\sum_{
  \left\{
    z_{+}
  \right\}}
  \left.
    \frac{e^{\mathrm{i}zL}}{(\mathcal{E}_{+}-\mathcal{E}_{-})\frac{\mathrm{d}\mathcal{E}_{+}}{\mathrm{d}z}}
  \right|_{z=z_{+}},
\end{equation}
where $\left\{ z_{+}(\rho)\right\} $ are the roots of the equation
$E=\mathcal{E}_{+}$. Since $L>0$, we close the contour in the upper
half-plane. All roots $\left\{ z_{+}\right\} $ are in the upper half-plane
as well.

In general case, the equation $E=\mathcal{E}_{+}$ is equivalent to
the algebraic equation of the 8th degree. However, if neutrinos are
ultrarelativistic, i.e. $E\gg m_{1,2}$, we can solve it perturbatively.
First, we represent Eq.~(\ref{eq:Epm}) in the form,
\begin{equation}\label{eq:Epultrel}
  \mathcal{E}_{+}\approx q+\sqrt{
  \left(
    \frac{\Delta m^{2}}{4q}+\frac{g_{2}-g_{1}}{2}
  \right)^{2}
  +g^{2}},
\end{equation}
where $\Delta m^{2}=m_{2}^{2}-m_{1}^{2}$. In the first approximation,
we take $q(z_{+})=\sqrt{z_{+}^{2}+\rho^{2}}\approx\mathcal{E}_{+}(E)$,
where
\begin{equation}\label{eq:Ep(E)}
  \mathcal{E}_{+}(E)=E+\sqrt{
  \left(
    \frac{\Delta m^{2}}{4E}+\frac{g_{2}-g_{1}}{2}
  \right)^{2}
  +g^{2}}.
\end{equation}
Analogously to Eq.~(\ref{eq:Ep(E)}), one finds the expression for
$\mathcal{E}_{-}(E)$. Finally, we get that the poles are $z_{+}=\sqrt{\mathcal{E}_{+}^{2}(E)-\rho^{2}}+\mathrm{i}0$
if $0<\rho<\mathcal{E}_{+}(E)$, and $z_{+}=\mathrm{i}\sqrt{\rho^{2}-\mathcal{E}_{+}^{2}(E)}$
if $\mathcal{E}_{+}(E)<\rho<\infty$. After the integrations, the
expression for $I$ takes the form,
\begin{equation}\label{eq:Iultrel}
  I=-\frac{Ee^{\mathrm{i}\mathcal{E}_{+}(E)L}}{2\pi L\sqrt{
  \left(
    \frac{\Delta m^{2}}{4E}+\frac{g_{2}-g_{1}}{2}
  \right)^{2}
  +g^{2}}}.
\end{equation}
The rest of the integrals contributing to the matrix element in Eq.~(\ref{eq:Memu})
is computed analogously.

We present the final result for the matrix element in Eq.~(\ref{eq:Memu}),
\begin{align}\label{eq:matrelraw}
  \mathcal{M}_{e\to\mu}= & -\frac{
  \left(
    \frac{\Delta m^{2}}{4E}+\frac{g_{2}-g_{1}}{2}
  \right)
  \sin2\theta+g\cos2\theta}{\sqrt{
  \left(
    \frac{\Delta m^{2}}{4E}+\frac{g_{2}-g_{1}}{2}
  \right)^{2}
  +g^{2}}}\sin
  \left(
    \sqrt{
    \left(
      \frac{\Delta m^{2}}{4E}+\frac{g_{2}-g_{1}}{2}
    \right)^{2}
    +g^{2}|}L|
  \right)
  \nonumber
  \\
  & \times
  \frac{\mathrm{i}Ee^{\mathrm{i}E|L|}}{2\pi|L|}\kappa_{-}^{\dagger}(p_{\mu})
  \left[
    1-(\bm{\sigma}\hat{L})
  \right]
  \kappa_{-}(p_{e}).
\end{align}
Analogously to Appendix~\ref{sec:FOURTRANS}, we assume that the
neutrino propagation distance is large in deriving of Eq.~(\ref{eq:matrelraw}).
If the incoming and outgoing leptons are ultrarelativistic and their
momenta are almost along the same direction, the quantity $\kappa_{-}^{\dagger}(p_{\mu})\left[1-(\bm{\sigma}\hat{L})\right]\kappa_{-}(p_{e})$
is just a nonzero constant factor.

The effective potentials of the mass eigenstates interaction with
matter are
\begin{equation}
  g_{1}=V_{e}\cos^{2}\theta+V_{\mu}\sin^{2}\theta,
  \quad
  g_{2}=V_{e}\sin^{2}\theta+V_{\mu}\cos^{2}\theta,
  \quad
  g=(V_{e}-V_{\mu})\sin\theta\cos\theta,
\end{equation}
where $V_{e,\mu}$ are given in Eq.~(\ref{eq:VeVmu}). Finally, we get that the
probability of the process $e\to\mu$ contains the factor,
\begin{align}\label{eq:MSWtrpr}
  P_{e\to\mu} = & \frac{
  \left(
    \frac{\Delta m^{2}}{4E}\sin2\theta
  \right)^{2}}{
  \left(
    \frac{\Delta m^{2}}{4E}\sin2\theta
  \right)^{2}+
  \left(
    \frac{\Delta m^{2}}{4E}\cos2\theta-\frac{G_{\mathrm{F}}n_{e}}{\sqrt{2}}
  \right)^{2}}
  \nonumber
  \\
  & \times\sin^{2}
  \left(\sqrt{
    \left(
      \frac{\Delta m^{2}}{4E}\sin2\theta
    \right)^{2}+
    \left(
      \frac{\Delta m^{2}}{4E}\cos2\theta-\frac{G_{\mathrm{F}}n_{e}}{\sqrt{2}}
    \right)^{2}}|L|
  \right),
\end{align}
which coincides with the prediction of the standard quantum mechanical treatment for neutrino flavor oscillations in uniform background matter.

\subsection{Comparison with the quantum mechanical approach\label{sec:QMvsQFT}}

Originally, in Ref.~\cite{Wol78}, the neutrino propagation and oscillations in matter were studied within the quantum mechanical approach. First, we discuss the case of a uniform matter. Considering two flavor eigenstates, $\nu_f^\mathrm{T} = (\nu_{e},\nu_{\mu})$, one gets that their evolution obeys the effective Schr\"odinger equation~\cite{MohPal04},
\begin{equation}\label{eq:Scheqeff}
  \mathrm{i}\frac{\mathrm{d}\nu_f}{\mathrm{d}x} =
  H_\mathrm{eff} \nu_f,
  \quad
  H_\mathrm{eff} =
  \left(
    \begin{array}{cc}
      -\frac{\Delta m^{2}}{4E}\cos2\theta + \sqrt{2}G_{\mathrm{F}}n_{e} & \frac{\Delta m^{2}}{4E}\sin2\theta \\
     \frac{\Delta m^{2}}{4E}\sin2\theta & \frac{\Delta m^{2}}{4E}\cos2\theta
    \end{array}
  \right),
\end{equation}
where we assume that the beam of ultrarelativistic neutrinos propagates in space, i.e. $x\approx t$.

If $n_e$ in Eq.~\eqref{eq:Scheqeff} does not depend on coordinates, the transition probability for the process $\nu_e \to \nu_\mu$, based on Eq.~\eqref{eq:Scheqeff}, coincides with that in Eq.~\eqref{eq:MSWtrpr} when $x=L$. Now, we assume that the matter density changes adiabatically from a great value to zero. In this case, we can split the trajectory into thin layers, with the density in any of them being approximately constant. If a neutrino beam travels through such a matter, it will reach the layer with the density $n_e^{(\mathrm{res})} = \tfrac{\Delta m^{2} \cos2\theta}{2\sqrt{2}G_{\mathrm{F}}E}$, where the transition probability significantly increases. This phenomenon is called the resonant amplification of neutrino oscillations in matter, or the MSW effect.

The consideration of coordinate dependent matter density is problematic with the QFT approach studied here. Firstly, the propagators, obtained in Appendix~\ref{sec:MATTPROP}, have quite complicated form if $g_a = g_a(\mathbf{x})$. Second, the QFT formalism, as it was developed in Refs.~\cite{Kob82,GriSto96}, implies the derivation of probabilities rather than an evolution equation, like in Eq.~\eqref{eq:Scheqeff}. It is impossible to get the analytical transition probability for the arbitrarily dependent matter density. Nevertheless, we can consider the expression in Eq.~\eqref{eq:MSWtrpr} as the probability within a thin layer with approximately constant density provided the adiabaticity condition is fulfilled~\cite{MohPal04}.

The relativistic quantum mechanics approach, where neutrinos are treated as first quantized fields, was developed in Refs.~\cite{DvoMaa07,DvoMaa09,Dvo12}. It allows one to rederive the effective Schr\"odinger equation for neutrino oscillations in various external fields using the exact solutions of the Dirac equation accounting for these external fields. Hence, in that approach, one can consider time and coordinate dependent external fields (see, e.g., Refs.~\cite{Dvo18,Dvo19}). However, as mentioned in Sec.~\ref{sec:INTRO}, the relativistic quantum mechanics description of neutrino oscillations has a shortcoming of a quite broad initial neutrino wavepacket.

\section{Conclusion\label{sec:CONCL}}

In the present work, we have studied neutrino flavor oscillations
in background matter in frames of the QFT approach where massive neutrinos
are virtual particles. We have started in Sec.~\ref{sec:QFTFORMALISM}
with reminding of the basic concepts of this method. The calculation
of the matrix element for the process corresponding to oscillating
neutrinos involves propagators of mass eigenstates. We have established
in Sec.~\ref{sec:NUMATT} that mass eigenstates are mixed in background
matter in general case. It was the main difficulty in the application
of QFT for neutrino oscillations in matter.

In Sec.~\ref{sec:EXPROP}, we have derived the equations for the
exact propagators, which account for the neutrino interaction with
matter. These equations are analogous to the Dyson equations. Moreover,
we have established that the developed formalism is not applicable
for the Dirac mass eigenstates. For this reason, in Sec.~\ref{sec:REFORMMAJ},
we have reformulated the basic concepts of our approach in terms of
Weyl spinors, which are the representation of Majorana neutrinos.
The exact propagators for this kind of neutrinos have been found in
Sec.~\ref{sec:PROPWEYLNU}. Despite sizable experimental efforts are made (see, e.g., Ref.~\cite{Ago24}), the issue whether neutrinos are Dirac or Majorana particles is still open. Nevertheless, Majorana neutrinos are preferable from the theoretical point of view since they provide a natural explanation of small masses of active neutrinos~\cite{MirVal16}.

In Sec.~\ref{sec:MSW}, we have applied our findings to a particular
case of $\nu_{e}\to\nu_{\mu}$ oscillations. We have derived the matrix
element and the corresponding transition probability. One can see
in Eq.~(\ref{eq:MSWtrpr}) that it reproduces the transition probability for flavor oscillations in matter. We have compared our QFT based approach with the standard quantum mechanical description of neutrino oscillations in Sec.~\ref{sec:QMvsQFT}.

The diagonal propagators of Weyl neutrinos have been found in Appendix~\ref{sec:MATTPROP}.
The specific 3D Fourier transforms of the propagators have been computed
in Appendix~\ref{sec:FOURTRANS}.

Despite it is difficult to get something beyond Eq.~(\ref{eq:MSWtrpr})
which was originally derived within the quantum mechanical approach,
it is instructive to apply the exact formalism of QFT for neutrino
flavor oscillations mainly from the methodological point of view.
It allows one to analyze the validity of approximations made to obtain
Eq.~(\ref{eq:MSWtrpr}). The main assumptions used in our work are
(i) that neutrinos are ultrarelativistic particles; and (ii) the weakness
of the matter interaction.

The assumption (i) was used first in Sec.~\ref{sec:REFORMMAJ}, where
it allowed us to exclude the involvement of antineutrinos in the series
in Eqs.~(\ref{eq:Sigmaaser}) and~(\ref{eq:Sigma12ser}). Without
this assumption, the summation of the diagrams in Fig.~\ref{fig:feynQFT}
would be quite nontrivial. Second, the assumption (i) was utilized
in Appendix~\ref{sec:FOURTRANS} in the computation of the integrals
$I_{i}$ and $J_{i}$. It allowed us to show that the contribution
of $I_{1}$ and $J_{1}$ is the leading one. Thus, we could reduce
the number of terms in $S_{a}$; cf. Eq.~(\ref{eq:ABultrarel}).
Third, the assumption (i) was used in the calculation of the 3D Fourier
transform in Eqs.~(\ref{eq:Epultrel})-(\ref{eq:Iultrel}) resulting
in the matrix element in Eq.~(\ref{eq:matrelraw}). Thus, we can
see that the ultrarelativity of neutrinos is crucial for Eq.~(\ref{eq:MSWtrpr}).

The assumption (ii) was used to simplify the propagators and jump
from Eq.~(\ref{eq:SigmaAB}) to Eq.~(\ref{eq:Sigmaab2fl}). In fact,
it allowed us to get the $g^{2}$-term under the square root in Eq.~(\ref{eq:Epm}).
In principle, one can study corrections to Eq.~(\ref{eq:MSWtrpr})
for neutrinos propagating in extremely dense matter.

There are a couple of additional important assumptions used in our work. The original formulation of the QFT method in Refs.~\cite{Kob82,GriSto96} implied that one gets the probabilities rather than the evolution equation. That is why, we assume that the matter density is constant. The discussion of the possibility to take into account a nonuniform matter has been provided in Sec.~\ref{sec:QMvsQFT}.

We have also studied the system of two neutrinos. It allowed us to derive the analytical transition probability in Eq.~\eqref{eq:MSWtrpr}. One can notice in Fig.~\ref{fig:DysonFeyn} that the matter interaction induces the transitions like $1\to 2$ or $2\to 1$. It makes possible to sum the series in Eqs.~\eqref{eq:Sigmaaser} and~\eqref{eq:Sigma12ser}. If more mass eigenstates were considered, it would be difficult to derive general Eqs.~\eqref{eq:DysonSigmaa} and~\eqref{eq:DysonSigma12} since the Feynman graphs would include greater number of transitions.

An additional reason why oscillations of Dirac neutrinos cannot be studied within the described approach results from the aforementioned argument. Indeed, in case of Dirac neutrinos, one has to consider more than two equivalent Majorana degrees of freedom. In other words, for Dirac neutrinos, we have to sum the infinite number of terms, like in Eqs.~\eqref{eq:Sigmaaser} and~\eqref{eq:Sigma12ser}, each of them being a branching diagram. This task seems to be quite challenging.

\appendix

\section{Massive Weyl neutrino propagators in matter\label{sec:MATTPROP}}

In this Appendix, we derive the propagators of massive Weyl neutrinos
in matter. The Lagrangian for a single mass eigenstate, described
by the two component wave function $\eta_{a}$, having the mass $m_{a}$
and interacting with nonmoving and unpolarized matter is
\begin{equation}\label{eq:WeylLagr}
  \mathcal{L}=\mathrm{i}\eta_{a}^{\dagger}(\sigma^{\mu}\partial_{\mu})\eta_{a}-g\eta_{a}^{\dagger}\eta_{a}
  +\frac{\mathrm{i}}{2}m_{a}\eta_{a}^{\dagger}\sigma_{2}\eta_{a}^{*}-\frac{\mathrm{i}}{2}m\eta_{a}^{\mathrm{T}}\sigma_{2}\eta_{a},
\end{equation}
where $\sigma^{\mu}=(1,-\bm{\sigma})$, and $g_{a}$ is given in Sec.~\ref{sec:NUMATT}.
Based on Eq.~(\ref{eq:WeylLagr}), we obtain the wave equation for
$\eta_{a}$
\begin{equation}\label{eq:weqWeyl}
  \mathrm{i}\dot{\eta}_{a}+[(\bm{\sigma}\mathbf{p})-g_{a}]\eta_{a}+\mathrm{i}m_{a}\sigma_{2}\eta_{a}^{*}=0.
\end{equation}
Analogous equation is obtained for $\eta_{a}^{*}$.

The general solution of Eq.~(\ref{eq:weqWeyl}) is
\begin{align}\label{eq:Weylsol}
  \eta_{a}(\mathbf{x},t)= & \int\frac{\mathrm{d}^{3}p}{(2\pi)^{3/2}}
  \bigg[
    \lambda_{a-}
    \left(
      a_{-}w_{-}e^{-\mathrm{i}E_{a-}t+i\mathbf{px}}+A_{a-}a_{-}^{\dagger}w_{+}e^{\mathrm{i}E_{a-}t-\mathrm{i}\mathbf{px}}
    \right)
    \nonumber
    \\
    & +
    \lambda_{a+}
    \left(
      A_{a+}a_{+}w_{+}e^{-\mathrm{i}E_{a+}t+\mathrm{i}\mathbf{px}}+a_{+}^{\dagger}w_{-}e^{\mathrm{i}E_{a+}t-\mathrm{i}\mathbf{px}}
    \right)
  \bigg],
\end{align}
where $E_{a\pm}=\sqrt{m_{a}^{2}+(p\mp g_{a})^{2}},$ are the energies
for right and left polarized neutrinos, $a_{\pm}^{\dagger}$ and $a_{\pm}$
are the creation and annihilation operators for certain helicities,
\begin{equation}\label{eq:lamA}
  \lambda_{a\pm}^{2}=\frac{E_{a\pm}+p\mp g_{a}}{2E_{a\pm}},
  \quad
  A_{a\pm}=\mp\frac{m_{a}}{E_{a\pm}+p\mp g_{a}},
\end{equation}
and
\begin{equation}\label{eq:helamp}
  w_{+}=
  \left(
    \begin{array}{c}
      e^{-\mathrm{i}\phi/2}\cos\vartheta/2
      \\
      e^{\mathrm{i}\phi/2}\sin\vartheta/2
    \end{array}
  \right),
  \quad
  w_{-}=
  \left(
    \begin{array}{c}
      -e^{-\mathrm{i}\phi/2}\sin\vartheta/2
      \\
      e^{\mathrm{i}\phi/2}\cos\vartheta/2
    \end{array}
  \right),
\end{equation}
are the helicity amplitudes. In Eq.~(\ref{eq:helamp}), the angles
$\vartheta$ and $\phi$ fix the direction of the neutrino momentum
$\mathbf{p}$. We mention some important properties of $w_{\pm}$,
\begin{equation}\label{eq:helampprop}
  w_{\pm}\otimes w_{\pm}^{\dagger}=\frac{1}{2}
  \left[
    1\pm\bm{\sigma}\hat{p}
  \right],
  \quad
  w_{\pm}\otimes w_{\mp}^{\mathrm{T}}=\pm\frac{1}{2}
  \left(
    1\pm\bm{\sigma}\hat{p}
  \right)
  \mathrm{i}\sigma_{2},
\end{equation}
which can be checked by means direct calculations using Eq.~(\ref{eq:helamp}).

Note that $\eta_{a}(\mathbf{x},t)$ in Eq.~(\ref{eq:Weylsol}) and
analogous $\eta_{a}^{*}(\mathbf{x},t)$ obey the canonical equal time
anticommutation relation $\left\{ \eta_{a}(\mathbf{x},t),\eta_{a}^{*}(\mathbf{y},t)\right\} =\delta(\mathbf{x}-\mathbf{y})$
provided that $\left\{ a_{\pm}(\mathbf{p}),a_{\pm}^{\dagger}(\mathbf{q})\right\} =\delta(\mathbf{p}-\mathbf{q})$
and all other anticommutators vanish. Thus, we have constructed the
canonical quantization of Weyl neutrinos in background matter.

We define two propagators for Weyl neutrinos
\begin{align}
  S_{a}(x-y) = &
  \left\langle
    0
    \left|
      T[\eta_{a}(x)\eta_{a}^{\dagger}(y)]
    \right|
    0
  \right\rangle
  = \theta(x_{0}-y_{0})
  \left\langle
    0
    \left|
      \eta_{a}(x)\eta_{a}^{\dagger}(y)
    \right|
    0
  \right\rangle
  \notag
  \\
  & -
  \theta(y_{0}-x_{0})
  \left\langle
    0
    \left|
      \eta_{a}^{*}(y)\eta_{a}^{\mathrm{T}}(x)
    \right|
    0
  \right\rangle,
\end{align}
and
\begin{align}
  \tilde{S}_{a}(x-y) = &
  \left\langle
    0
    \left|
      T[\eta_{a}(x)\eta_{a}^{\mathrm{T}}(y)]
    \right|
    0
  \right\rangle
  = \theta(x_{0}-y_{0})
  \left\langle
    0
    \left|
      \eta_{a}(x)\eta_{a}^{\mathrm{T}}(y)
    \right|
    0
  \right\rangle
  \notag
  \\
  & -
  \theta(y_{0}-x_{0})
  \left\langle
    0
    \left|
      \eta_{a}(y)\eta_{a}^{\mathrm{T}}(x)
    \right|
    0
  \right\rangle ,
\end{align}
where $\theta(t)$ is the Heaviside step function. Using Eqs.~(\ref{eq:Weylsol}),
(\ref{eq:lamA}), and~(\ref{eq:helampprop}), as well as the identity
\begin{equation}
  e^{\pm\mathrm{i}Et}\theta(\mp t)=
  \pm\frac{1}{2\pi\mathrm{i}}\int_{-\infty}^{+\infty}\frac{e^{-\mathrm{i}p_{0}t}\mathrm{d}p_{0}}{p_{0}\pm E\mp\mathrm{i}0},
\end{equation}
where $\mathrm{i}0$ means a small imaginary part, we get the propagators
in the form,
\begin{align}
  S_{a}(x)= & \frac{\mathrm{i}}{2}\int\frac{\mathrm{d}^{4}p}{(2\pi)^{4}}e^{-\mathrm{i}px}
  \bigg\{
    \bigg[
      \lambda_{a-}^{2}
      \left(
        \frac{1}{p_{0}-E_{a-}+\mathrm{i}0}+\frac{A_{a-}^{2}}{p_{0}+E_{a-}-\mathrm{i}0}
      \right)
      \notag
      \displaybreak[2]
      \\
      & +
      \lambda_{a+}^{2}
      \left(
        \frac{1}{p_{0}+E_{a+}-\mathrm{i}0}+\frac{A_{a+}^{2}}{p_{0}-E_{a+}+\mathrm{i}0}
      \right)
    \bigg]
    \nonumber
    \displaybreak[2]
    \\
    & -
    (\bm{\sigma}\hat{p})
    \bigg[
      \lambda_{a-}^{2}
      \left(
        \frac{1}{p_{0}-E_{a-}+\mathrm{i}0}+\frac{A_{a-}^{2}}{p_{0}+E_{a-}-\mathrm{i}0}
      \right)
      \notag
      \displaybreak[2]
      \\
      & -
      \lambda_{a+}^{2}
      \left(
        \frac{1}{p_{0}+E_{a+}-\mathrm{i}0}+\frac{A_{a+}^{2}}{p_{0}-E_{a+}+\mathrm{i}0}
      \right)
    \bigg]
  \bigg\},
  \label{eq:Sprop}
  \displaybreak[2]
  \\
  \tilde{S}_{a}(x)= & \frac{m\sigma_{2}}{2}\int\frac{\mathrm{d}^{4}p}{(2\pi)^{4}}e^{-\mathrm{i}px}
  \left[
    \frac{
    \left(
      1-\bm{\sigma}\hat{p}
    \right)
    }{p_{0}^{2}-E_{a-}^{2}+\mathrm{i}0}+\frac{
    \left(
      1+\bm{\sigma}\hat{p}
    \right)}{p_{0}^{2}-E_{a+}^{2}+\mathrm{i}0}
  \right].
  \label{eq:tildeSprop}
\end{align}
Note that, if one turns off the neutrino interaction with matter,
then $E_{a+}=E_{a-}=\sqrt{p^{2}+m_{a}^{2}}$. Thus, $S_{a}$ and $\tilde{S}_{a}$
in Eqs.~(\ref{eq:Sprop}) and~(\ref{eq:tildeSprop}) reproduce the
known vacuum propagators for a massive Weyl neutrino in Ref.~\cite{FukYan03}.

\section{3D Fourier transform of propagators\label{sec:FOURTRANS}}

In this Appendix, we calculate the Fourier transform of the propagator
of a single massive Weyl neutrino in matter. The corresponding
propagators are taken from Appendix~\ref{sec:MATTPROP}.

In Sec.~\ref{sec:QFTFORMALISM}, we obtain that the matrix element
contains the Fourier transform of the propagator in the form,
\begin{equation}\label{eq:ftprop}
  \int\frac{\mathrm{d}^{3}q}{(2\pi)^{3}}S_{a}(E,\mathbf{q})e^{\mathrm{i}\mathbf{qL}},
\end{equation}
where $E$ is the mean energy of incoming and outgoing leptons and
$\mathbf{L}$ is the vector between the source and the detector. One
can see in Eq.~(\ref{eq:Sprop}) that the Fourier image of the propagator
contains eight terms. We provide the detailed calculations only for
two of them,
\begin{equation}\label{eq:I1J1}
  I_{1}=\int\frac{\mathrm{d}^{3}q}{(2\pi)^{3}}\frac{\lambda_{a-}^{2}e^{\mathrm{i}\mathbf{qL}}}{E-E_{a-}+\mathrm{i}0},
  \quad
  J_{1}=\int\frac{\mathrm{d}^{3}q}{(2\pi)^{3}}\frac{\lambda_{a-}^{2}(\bm{\sigma}\hat{q})e^{\mathrm{i}\mathbf{qL}}}{E-E_{a-}+\mathrm{i}0}.
\end{equation}
The rest of the integrals is computed analogously.

We use the cylindrical coordinates $\mathbf{q}=(\rho,z,\phi)$ and
suppose that $\mathbf{L}\parallel\mathbf{e}_{z}$. For example,
\begin{equation}\label{eq:I1-1}
  I_{1}=\frac{1}{(2\pi)^{2}}\int_{0}^{+\infty}\rho\mathrm{d}\rho\int_{-\infty}^{+\infty}\mathrm{d}z
  \frac{(E_{a-}+q+g_{a})e^{\mathrm{i}zL}}{2E_{a-}(E-\sqrt{m_{a}^{2}+(\sqrt{z^{2}+\rho^{2}}+g_{a})^{2}}+\mathrm{i}0)}.
\end{equation}
In the integrand in Eq.~(\ref{eq:I1-1}), we take into account that
$E_{a-}=\sqrt{m_{a}^{2}+(\sqrt{z^{2}+\rho^{2}}+g_{a})^{2}}$ and $q=\sqrt{z^{2}+\rho^{2}}$.

We expand the function $Z(z)=E-\sqrt{m_{a}^{2}+(\sqrt{z^{2}+\rho^{2}}+g_{a})^{2}}$
in series near the pole $z_{0}$, $Z(z_{0})=0$,
\begin{equation}
  Z\approx Z'(z-z_{0})+\dotsc,
  \quad
  Z'=\frac{\mathrm{d}}{\mathrm{d}z}
  \left[
    E-\sqrt{m_{a}^{2}+(\sqrt{z^{2}+\rho^{2}}+g_{a})^{2}}
  \right]_{z=z_{0}}=
  -\frac{z_{0}(q_{0}+g_{a})}{q_{0}E},
\end{equation}
where $q_{0}=q(z_{0})=\sqrt{z_{0}^{2}+\rho^{2}}$. We have two cases:
(i) $z_{0}=\pm\sqrt{\rho_{0}^{2}-\rho^{2}}\pm\mathrm{i}0$, when $\rho<\rho_{0}$;
and (ii) $z_{0}=\pm\mathrm{i}\sqrt{\rho^{2}-\rho_{0}^{2}}$, when
$\rho>\rho_{0}$. Here $\rho_{0}=\sqrt{E^{2}-m_{a}^{2}}-g_{a}$. Note
that $q_{0}=\rho_{0}$. The imaginary parts of residues in the $z$-variable
are synchronized with that in Eq.~(\ref{eq:I1-1}).

If $L>0$, we close the contour in the upper half-plane in both cases
while integrating over the $z$-variable. In this situation, the residues
contributing to the integration are $z_{0}=\sqrt{\rho_{0}^{2}-\rho^{2}}+\mathrm{i}0$
in case (i) and $z_{0}=\mathrm{i}\sqrt{\rho^{2}-\rho_{0}^{2}}$ in
case (ii). Splitting the integration over $\rho$ into two parts,
$0<\rho<\rho_{0}$ and $\rho_{0}<\rho<\infty$, one has that
\begin{align}\label{eq:I1-2}
  I_{1}= & -\frac{\mathrm{i}\rho_{0}(E+\rho_{0}+g_{a})}{4\pi(\rho_{0}+g_{a})}
  \left(
    \int_{0}^{\rho_{0}}\rho\mathrm{d}\rho\frac{e^{\mathrm{i}\sqrt{\rho_{0}^{2}-\rho^{2}}L}}{\sqrt{\rho_{0}^{2}-\rho^{2}}}
    -\mathrm{i}\int_{\rho_{0}}^{\infty}\rho\mathrm{d}\rho\frac{e^{-\sqrt{\rho^{2}-\rho_{0}^{2}}L}}{\sqrt{\rho^{2}-\rho_{0}^{2}}}
  \right)
  \nonumber
  \\
  & =
  -\frac{
  \left(
    E+\sqrt{E^{2}-m_{a}^{2}}
  \right)
  \left(
    \sqrt{E^{2}-m_{a}^{2}}-g
  \right)
  }{4\pi L\sqrt{E^{2}-m_{a}^{2}}}e^{\mathrm{i}(\sqrt{E^{2}-m_{a}^{2}}-g_{a})L}.
\end{align}
If $L<0$, we close the contour in the lower half-plane and the residues
are $z_{0}=-\sqrt{\rho_{0}^{2}-\rho^{2}}-\mathrm{i}0$ in case (i)
and $z_{0}=-\mathrm{i}\sqrt{\rho^{2}-\rho_{0}^{2}}$ in case (ii).
The integration over $\rho$ is analogous to that in Eq.~(\ref{eq:I1-2}).
The final result is
\begin{align}\label{eq:I1}
  I_{1} & =-\frac{
  \left(
    E+\sqrt{E^{2}-m_{a}^{2}}
  \right)
  \left(
    \sqrt{E^{2}-m_{a}^{2}}-g_{a}
  \right)
  }{4\pi|L|\sqrt{E^{2}-m_{a}^{2}}}e^{\mathrm{i}(\sqrt{E^{2}-m_{a}^{2}}-g_{a})|L|},
\end{align}
where we incorporate both positive and negative values of $L$.

The integrand in $J_{1}$ in Eq.~(\ref{eq:I1J1}) contains the matrix
factor
\begin{equation}
  (\bm{\sigma}\hat{q})=\frac{1}{\sqrt{z^{2}+\rho^{2}}}
  \left(
    \begin{array}{cc}
      z & \rho e^{-\mathrm{i}\phi}
      \\
      \rho e^{\mathrm{i}\phi} & -z
    \end{array}
  \right),
\end{equation}
in addition to that in Eq.~(\ref{eq:I1-1}). The integration over
$\phi$, analogous to that in Eq.~(\ref{eq:I1-1}), removes the off-diagonal
terms in the final result. Thus, we should calculate only the diagonal
elements which differ from these in Eq.~(\ref{eq:I1-1}) by the additional
factors $\pm z/\sqrt{z^{2}+\rho^{2}}$ in the integrands. Since the
computation is almost identical to that for $I_{1}$, we just present
the final result,
\begin{multline}
  \int\frac{\mathrm{d}^{3}q}{(2\pi)^{3}}\frac{\lambda_{a-}^{2}e^{\mathrm{i}\mathbf{qL}}}{E-E_{a-}+\mathrm{i}0}
  \left(
    \pm\frac{z}{\sqrt{z^{2}+\rho^{2}}}
  \right)
  = 
  \mp\frac{
  \left(
    E+\sqrt{E^{2}-m_{a}^{2}}
  \right)
  \left(
    \sqrt{E^{2}-m_{a}^{2}}-g_{a}
  \right)
  }{4\pi L\sqrt{E^{2}-m_{a}^{2}}}
  \\
  \times
  e^{\mathrm{i}(\sqrt{E^{2}-m_{a}^{2}}-g_{a})|L|}
  \left(
    1+\frac{\mathrm{i}}{|L|(\sqrt{E^{2}-m_{a}^{2}}-g_{a})}
  \right),
\end{multline}
which is valid for arbitrary values of $L$. Taking into account the
chosen direction of $\mathbf{L}$, the final result for $J_{1}$ reads
\begin{align}\label{eq:J1}
  J_{1}= & \int\frac{\mathrm{d}^{3}q}{(2\pi)^{3}}\frac{\lambda_{a-}^{2}(\bm{\sigma}\hat{q})e^{\mathrm{i}\mathbf{qL}}}{E-E_{a-}+\mathrm{i}0}=
  -(\bm{\sigma}\hat{L})\frac{
  \left(
    E+\sqrt{E^{2}-m_{a}^{2}}
  \right)
  \left(
    \sqrt{E^{2}-m_{a}^{2}}-g_{a}
  \right)
  }{4\pi|L|\sqrt{E^{2}-m_{a}^{2}}}
  \nonumber
  \\
  & \times
  e^{\mathrm{i}(\sqrt{E^{2}-m_{a}^{2}}-g_{a})|L|}
  \left(
    1+\frac{\mathrm{i}}{|L|(\sqrt{E^{2}-m_{a}^{2}}-g_{a})}
  \right).
\end{align}
Considering the limit of large neutrino propagation distance,
one has that $J_{1}\approx(\bm{\sigma}\hat{L})I_{1}$.

Eventually, we list the results for the rest of the terms in Eq.~(\ref{eq:Sprop}).
They are
\begin{align}\label{eq:I2-4}
  I_{2} & =\int\frac{\mathrm{d}^{3}q}{(2\pi)^{3}}\frac{\lambda_{a+}^{2}A_{a+}^{2}e^{\mathrm{i}\mathbf{qL}}}{E-E_{a+}+\mathrm{i}0}=
  -\frac{m_{a}^{2}
  \left(
    \sqrt{E^{2}-m_{a}^{2}}+g_{a}
  \right)
  }{4\pi|L|
  \left(
    E+\sqrt{E^{2}-m_{a}^{2}}
  \right)
  \sqrt{E^{2}-m_{a}^{2}}}
  \notag
  \displaybreak[2]
  \\
  & \times
  e^{\mathrm{i}(\sqrt{E^{2}-m_{a}^{2}}+g_{a})|L|},
  \nonumber
  \displaybreak[2]
  \\
  I_{3} & =\int\frac{\mathrm{d}^{3}q}{(2\pi)^{3}}\frac{\lambda_{a-}^{2}A_{a-}^{2}e^{\mathrm{i}\mathbf{qL}}}{E+E_{a-}-\mathrm{i}0}=
  -\frac{m_{a}^{2}
  \left(
    \sqrt{E^{2}-m_{a}^{2}}-g_{a}
  \right)
  }{4\pi|L|
  \left(
    E-\sqrt{E^{2}-m_{a}^{2}}
  \right)
  \sqrt{E^{2}-m_{a}^{2}}}
  \notag
  \displaybreak[2]
  \\
  & \times
  e^{-\mathrm{i}(\sqrt{E^{2}-m_{a}^{2}}-g_{a})|L|},
  \nonumber
  \displaybreak[2]
  \\
  I_{4} & =\int\frac{\mathrm{d}^{3}q}{(2\pi)^{3}}\frac{\lambda_{a+}^{2}e^{\mathrm{i}\mathbf{qL}}}{E+E_{a+}-\mathrm{i}0}=
  -\frac{
  \left(
    E-\sqrt{E^{2}-m_{a}^{2}}
  \right)
  \left(
    \sqrt{E^{2}-m_{a}^{2}}+g_{a}
  \right)
  }{4\pi|L|\sqrt{E^{2}-m_{a}^{2}}}
  \notag
  \displaybreak[2]
  \\
  & \times
  e^{-\mathrm{i}(\sqrt{E^{2}-m_{a}^{2}}+g_{a})|L|},
\end{align}
and
\begin{align}\label{eq:J2-4}
  J_{2}= & \int\frac{\mathrm{d}^{3}q}{(2\pi)^{3}}
  \frac{\lambda_{a+}^{2}A_{a+}^{2}(\bm{\sigma}\hat{q})e^{\mathrm{i}\mathbf{qL}}}{E-E_{a+}+\mathrm{i}0} \approx
  -(\bm{\sigma}\hat{L})\frac{m_{a}^{2}
  \left(
    \sqrt{E^{2}-m_{a}^{2}}+g_{a}\right)}{4\pi|L|\left(E+\sqrt{E^{2}-m_{a}^{2}}
  \right)
  \sqrt{E^{2}-m_{a}^{2}}}    
  \nonumber
  \\
  & \times
  e^{\mathrm{i}(\sqrt{E^{2}-m_{a}^{2}}+g_{a})|L|}=(\bm{\sigma}\hat{L})I_{2},
  \nonumber
  \\
  J_{3}= & \int\frac{\mathrm{d}^{3}q}{(2\pi)^{3}}
  \frac{\lambda_{a-}^{2}A_{a-}^{2}(\bm{\sigma}\hat{q})e^{\mathrm{i}\mathbf{qL}}}{E+E_{a-}-\mathrm{i}0}\approx
  (\bm{\sigma}\hat{L})\frac{m_{a}^{2}
  \left(
    \sqrt{E^{2}-m_{a}^{2}}-g\right)}{4\pi|L|\left(E-\sqrt{E^{2}-m_{a}^{2}}
  \right)
  \sqrt{E^{2}-m_{a}^{2}}}
  \nonumber
  \\
  & \times
  e^{-\mathrm{i}(\sqrt{E^{2}-m_{a}^{2}}-g_{a})|L|}=-(\bm{\sigma}\hat{L})I_{3},
  \nonumber
  \\
  J_{4}= & \int\frac{\mathrm{d}^{3}q}{(2\pi)^{3}}
  \frac{\lambda_{a+}^{2}(\bm{\sigma}\hat{q})e^{\mathrm{i}\mathbf{qL}}}{E+E_{a+}-\mathrm{i}0}\approx
  (\bm{\sigma}\hat{L})\frac{
  \left(
    E-\sqrt{E^{2}-m_{a}^{2}}
  \right)
  \left(
    \sqrt{E^{2}-m_{a}^{2}}+g_{a}
  \right)
  }{4\pi|L|\sqrt{E^{2}-m_{a}^{2}}}
  \nonumber
  \\
  & \times
  e^{-\mathrm{i}(\sqrt{E^{2}-m_{a}^{2}}+g_{a})|L|}=-(\bm{\sigma}\hat{L})I_{4}.
\end{align}
In Eq.~(\ref{eq:J2-4}), we show the limit of great $L$. One can
see in Eqs.~(\ref{eq:I1}) and~(\ref{eq:I2-4}), that $|I_{1,3}|\gg|I_{2,4}|$
for ultrarelativistic particles with $E\gg m_{a}$.

According to Eq.~(\ref{eq:matrel}), the matrix element is proportional
to the Fourier transform of the propagator in Eq.~(\ref{eq:ftprop})
averaged over the lepton wavefunctions. For simplicity, we neglect
the mixing matrix in Eq.~(\ref{eq:matrel}). Considering ultrarelativistic
left leptons in initial and final states with $\mathbf{p}\parallel\mathbf{L}$,
as well as using Eqs.~(\ref{eq:I1}), (\ref{eq:J1}), (\ref{eq:I2-4}),
and~(\ref{eq:J2-4}), one has that
\begin{align}
  \mathcal{M} = & \kappa_{-}^{\dagger}(\mathbf{p})
  \int\frac{\mathrm{d}^{3}q}{(2\pi)^{3}}S(E,\mathbf{q})e^{\mathrm{i}\mathbf{qL}}\kappa_{-}(\mathbf{p})
  \notag
  \\
  & \approx
  \frac{\mathrm{i}}{2}\kappa_{-}^{\dagger}(\mathbf{p})
  \left\{
    (I_{1}+I_{2})[1-(\bm{\sigma}\hat{L})]+(I_{3}+I_{4})[1+(\bm{\sigma}\hat{L})]
  \right\}
  \kappa_{-}(\mathbf{p})
  \nonumber
  \\
  & \approx
  \mathrm{i}I_{1}\approx-\frac{\mathrm{i}E}{4\pi|L|}e^{\mathrm{i}(\sqrt{E^{2}-m_{a}^{2}}-g_{a})|L|},
\end{align}
since $[1+(\bm{\sigma}\hat{L})]\kappa_{-}(\mathbf{p})=0$ and $[1-(\bm{\sigma}\hat{L})]\kappa_{-}(\mathbf{p})=2\kappa_{-}(\mathbf{p})$.

Thus, if one averages the matrix element over the left-polarized states
corresponding to ultrarelativistic leptons, the main contribution
results from the $I_{1}$ and $J_{1}$ terms in the neutrino propagator
in Eq.~(\ref{eq:Sprop}). The contribution of the propagator in Eq.~(\ref{eq:tildeSprop})
to the matrix element is suppressed by the factor $m_{a}/E$, which
is small for ultrarelativistic neutrinos. These are the main results
of this appendix.

\end{document}